\documentclass[aps,prb,twocolumn,superscriptaddress]{revtex4-1}

\usepackage{fancyhdr} 
\usepackage{color}
\usepackage{hyperref} 

\usepackage{pdfpages} 
\makeatletter
\AtBeginDocument{\let\LS@rot\@undefined}
\makeatother

\usepackage{verbatim} 
\usepackage{amsmath} 
\usepackage{amsthm} 
\usepackage{amssymb}	
\usepackage{graphicx} 

\let\baraccent=\= 
\renewcommand{\=}[1]{\stackrel{#1}{=}} 

\newcommand{\gae}{\lower 2pt \hbox{$\,
    \buildrel{\scriptstyle >}\over {\scriptstyle \sim}\,$}}
\newcommand{\lae}{\lower 2pt \hbox{$\,
    \buildrel{\scriptstyle <}\over {\scriptstyle \sim}\,$}}

\newcommand{\abs}[1]{\left| #1 \right|} 

\newcommand{\pd}[2]{\frac{\partial #1}{\partial #2}} 
 
\newcommand{\ket}[1]{\big| #1 \big\rangle} 

\newcommand{\expectation}[1]{\langle #1 \rangle}

\newcommand{\Z}{\mathbb{Z}}

\newcommand{\avg}[1]{\overline{#1}} 

\usepackage{xcolor}
\usepackage{ulem}
\usepackage{cancel}
\definecolor{violet}{rgb}{0.58, 0.0, 0.83}

\begin{document}

\title{Strong-Disorder Renormalization Group for Periodically Driven Systems} 

\author{William Berdanier}
\email[]{wberdanier@berkeley.edu}
\affiliation{Department of Physics, University of California, Berkeley, CA 94720, USA}

\author{Michael Kolodrubetz}
\affiliation{Department of Physics, The University of Texas at Dallas, Richardson, Texas 75080, USA}

\author{S. A. Parameswaran}
\affiliation{The Rudolf Peierls Centre for Theoretical Physics, Clarendon Laboratory, University of Oxford, Oxford OX1 3PU, UK}

\author{Romain Vasseur}
\affiliation{Department of Physics, University of Massachusetts, Amherst, Massachusetts 01003, USA}

\date{\today}

\begin{abstract}
Quenched randomness can lead to robust non-equilibrium phases of matter in periodically driven (Floquet) systems. Analyzing transitions between such dynamical phases requires a method capable of treating the twin complexities of disorder and discrete time-translation symmetry. We introduce a real-space renormalization group approach, asymptotically exact in the strong-disorder limit, and exemplify its use on the periodically driven interacting quantum Ising model. We analyze the universal physics near the critical lines and multicritical point of this model, and  demonstrate the robustness of our results to the inclusion of weak interactions.
\end{abstract}
\maketitle

\section{Introduction}

The study of periodically driven (Floquet) systems lies at the frontier of our understanding of non-equilibrium quantum physics. Despite the restriction to only {\it discrete} time translation symmetry and the attendant lack of full energy conservation, great strides have been made in recent years in understanding these inherently out-of-equilibrium systems. Results of these investigations include proposals to engineer exotic effective Hamiltonians~\cite{RevModPhys.89.011004,doi:10.1080/00018732.2015.1055918,PhysRevLett.116.205301,0953-4075-49-1-013001,PhysRevX.4.031027,PhysRevB.82.235114,Lindner:2011aa,PhysRevX.3.031005,1367-2630-17-12-125014,PhysRevB.79.081406,Gorg:2018aa,PhysRevLett.116.125301}, efforts to classify driven symmetry-protected topological phases~\cite{PhysRevB.92.125107,PhysRevB.93.245145,PhysRevB.93.245146,PhysRevB.93.201103,potter_classification_2016,PhysRevB.94.125105,PhysRevB.94.214203,PhysRevB.95.195128}, and the discovery of discrete time translation symmetry breaking -- dubbed ``time crystallinity''~\cite{pi-spin-glass,else_floquet_2016,PhysRevX.7.011026,PhysRevLett.118.030401,PhysRevB.96.115127,PhysRevLett.119.010602,Zhang:2017aa,Choi:2017aa,PhysRevB.97.184301,PhysRevLett.120.180603} --- among many others.  Floquet systems have thus been shown to host rich phase structures~\cite{khemani_prl_2016}, both extending fundamental concepts of equilibrium statistical mechanics into the non-equilibrium realm, and admitting new possibilities forbidden in equilibrium. 

Any discussion of the late-time limits of periodically driven closed quantum systems must address the issue of thermalization. Since such systems lack energy conservation, the expectation is that as energy is injected into them at periodic intervals, they will heat up to an infinite-temperature Gibbs state, characterized by a Floquet generalization of the eigenstate thermalization hypothesis (ETH)~\cite{PhysRevA.43.2046,PhysRevE.50.888,0034-4885-81-8-082001}.
The possibility of an exponentially long ``prethermal'' regime notwithstanding~\cite{KUWAHARA201696,PhysRevLett.115.256803}, the only known generic (i.e., not fine-tuned) exceptions to this scenario are systems that exhibit the phenomenon of many-body localization (MBL)~\cite{BASKO20061126,PhysRevB.75.155111,PalHuse,doi:10.1146/annurev-conmatphys-031214-014726,doi:10.1146/annurev-conmatphys-031214-014701,1742-5468-2016-6-064010,ALET2018,2018arXiv180411065A}. Such systems can  avoid heating and retain  a notion of distinct phases of matter even far away from thermal equilibrium~\cite{PhysRevB.88.014206,BahriMBLSPT,PhysRevB.89.144201}. The presence of quenched randomness therefore allows us to sharply define Floquet phases, and naturally leads to the possibility of transitions between them~\cite{khemani_prl_2016,PhysRevLett.114.140401,PONTE2015196,ABANIN20161,PhysRevLett.115.030402}.  

One of the most potent tools for studying one-dimensional random Hamiltonians is the real space renormalization group (RSRG)~\cite{PhysRevB.22.1305,PhysRevB.51.6411,PhysRevB.50.3799,PhysRevLett.69.534}. Though initially introduced as a technique for finding the ground states of random spin chains, these were subsequently extended to study excited states in MBL systems (RSRG-X)~\cite{PhysRevX.4.011052,PhysRevLett.112.217204}. Such RSRG-X techniques are not only a powerful (though approximate) way to obtain the excited states of interacting many-body systems at strong disorder, but also can be used to characterize the universal critical behavior near dynamical transitions between distinct MBL phases. Remarkably, RSRG often becomes asymptotically exact, since the effective disorder strength controlling the approach grows without bound under renormalization. It is therefore natural to examine whether these Floquet MBL systems might be amenable to a similar RSRG approach, improving our understanding of Floquet phases and the transitions possible between them.

In this work, we introduce a real-space renormalization group method for Floquet systems, which we dub ``Floquet RSRG," and apply it to understand the criticality of a canonical Floquet system: the driven Ising chain~\cite{khemani_prl_2016}. We derive a generalization of the Schrieffer-Wolff transformation to unitary operators which serves as our basic technical workhorse, allowing us to perturbatively construct a renormalized Floquet operator that captures the effective dynamics over ever-increasing length scales as the RG progresses. Working in the Majorana fermion language and tracking the flow of the couplings under renormalization, we identify the criticality in the model, including at the multicritical point. In contrast to previous works, our method is generic to {many periodically-driven systems\footnote{Formally, our method requires that the Floquet unitary be separable into a `strong' piece $F_0$ and `weak' piece $e^{iV}$, as described in more detail in the main text.}}: it only assumes the existence of a Floquet operator $F$, and furthermore is not limited to non-interacting Floquet-Ising~\cite{1742-5468-2017-7-073301} or discrete time crystal~\cite{PhysRevLett.118.030401} models. Even when applied to the driven Ising model, our method does not require all bond terms to correspond to near-0 quasienergy and all field terms to be near 0~\cite{1742-5468-2017-7-073301} or near $\pi$~\cite{PhysRevLett.118.030401}. Crucially, this allows us to analyze the proliferation of domain walls  between $0$ and $\pi$ regions, leading to a full description of the critical lines and multicritical point of the model. 
We find precise agreement with an earlier, intuitive, picture we proposed in terms of topological domain walls~\cite{Berdanier201805796}, {generalizing the picture proposed for quantum groundstates by Damle and Huse~\cite{PhysRevLett.89.277203}}. This work therefore provides an exact microscopic justification for results that may also be deduced on general grounds, thereby affirming their universality. {\it In toto}, these two perspectives give a rather complete description of criticality in a specific example of a periodically driven system, and in particular provide a concrete example of how an effective description in terms of domain walls arises from the underlying microscopic Floquet physics. 

The remainder of this paper is organized as follows. In Section~\ref{sec:theory} we review some basic aspects of Floquet theory, introduce the Floquet Schrieffer-Wolff transformation, and outline the general framework of Floquet RSRG. In Section~\ref{sec:Ising} we exemplify the use of this method on the driven Ising model, working first within the free-fermion limit. In Section~\ref{sec:ints} we show how Floquet RSRG can be used to argue analytically  for the irrelevance of interactions in the strong-disorder limit. Finally, we close with a summary of the work and a discussion of future applications of Floquet RSRG (Sec.~\ref{sec:discussion}).

\section{\label{sec:theory} Floquet RSRG via Floquet Schrieffer-Wolff Transformations}

Floquet systems are defined by a time-periodic Hamiltonian $H(t) = H(t+T)$. Equivalently, the Hamiltonian has a discrete time translation symmetry by the drive period $T$. As with Bloch's theorem for Hamiltonians with discrete spatial translation symmetry, the eigenstates of a time-periodic Hamiltonian must satisfy $\ket{\psi_\alpha(t)} = e^{-i E_\alpha t} \ket{\phi_\alpha(t)}$, where $\ket{\phi_\alpha(t+T)} = \ket{\phi_\alpha(t)}$ in units where $\hbar = 1$~\cite{PhysRev.138.B979,PhysRevA.7.2203}. A central object of interest is the single-period time evolution operator, 
\begin{equation}
F \equiv U(T) = \mathcal T { e}^{-i  \int_0^T dt H(t) } = { e}^{-i T H_F},
\end{equation}
where $\mathcal T$ denotes that the exponential is time-ordered, and  the last equality defines the so-called Floquet Hamiltonian. In general, the Floquet Hamiltonian is quite different from $H(t)$ at any $t$, and may in fact be non-local. 
Crucially, $H_F$ has eigenvalues that are constrained to lie on a circle, so $H_F = H_F + 2\pi/ T$. This in general eliminates the notion of a ground state,
requiring us to consider  the entire spectrum of  eigenstates of the Floquet operator $F$.

Our goal is to study such Floquet systems in one dimension in the presence of strong disorder, using a real-space renormalization group (RSRG) approach. Also termed the `strong-disorder renormalization group,' this method was initially introduced as a means of constructing the ground state~\cite{PhysRevB.22.1305,PhysRevB.51.6411,PhysRevB.50.3799,PhysRevLett.69.534} of one-dimensional Hamiltonians with quenched disorder, and later {extended (under the name `RSRG-X') to study} the full spectrum\cite{PhysRevX.4.011052} of such systems. The basic steps involved at any given stage of this RG scheme are (i) to identify the largest coupling in the effective Hamiltonian $H$ at that stage (which sets the characteristic energy scale); (ii) to `solve' the effective local problem $H_0$ defined by turning off all other couplings, assumed to be much weaker by virtue of the broad distribution of couplings; and (iii)  performing perturbation theory to determine the new couplings mediated by virtual fluctuations between eigenstates of $H_0$, thus defining a new effective Hamiltonian that can be fed back into step (i) in the next iteration. Tracking the flows of the distributions of couplings under this RG gives access to various quantities including correlation functions~\cite{PhysRevB.51.6411,PhysRevB.50.3799,PhysRevLett.69.534}, dynamical properties~\cite{PhysRevLett.84.3434,PhysRevB.63.134424}, and the entanglement entropy~\cite{PhysRevLett.93.260602}. Crucially, in many cases this procedure has an asymptotically self-consistent justification: to wit, initially broad distributions of the couplings broaden further with increasing iterations, indicating a flow to an `infinite randomness' fixed point where the RG procedure becomes exact.

At the heart of such RSRG methods is the Schrieffer-Wolff (SW) transformation~\cite{PhysRev.149.491}, a perturbative unitary rotation that eliminates off-diagonal elements of the Hamiltonian $H$ with respect to a ``strong'' piece $H_0$. In particular, one writes $H = H_0 + V$, where $\abs{V} \ll \abs{H_0}$, then seeks a unitary operator $e^{i S}$ such that $[e^{iS} H e^{-iS}, H_0] = 0$ to the desired order in $V$. This gives a self-consistent equation for $S$ to each order, which can be readily solved. Projecting onto an eigenstate subspace of the strong-coupling piece, one finds the overall energy shift and renormalized couplings between the remaining degrees of freedom. Projecting onto the lowest-energy eigenstate in each step then picks out the ground state of the problem, whereas projecting onto different subspaces besides the lowest-energy state allows access to excited states --- this being the key modification involved in RSRG-X.

In a similar spirit, let us imagine decomposing a Floquet operator $F$ into the product of a ``strong''{\footnote{Generically, since the eigenvalues of a unitary operator lie on a circle, perturbation theory on a unitary operator is not well defined as the eigenvalues are not well-ordered. We note that care must be taken in identifying a ``strong'' piece such that perturbation theory is well-controlled; this can fail, as it does at the naive decimation of domain walls in the main text.}} piece $F_0$ and a ``weak'' piece $V$, where $V \in \mathcal O(\lambda)$ with $\lambda$ a small parameter:
\begin{equation}
F = F_0 e^{i V}.
\end{equation}
The ordering of $F_0$ and $e^{i V}$ is an arbitrary choice of convention. We then seek a unitary operator $e^{i S}$ that transforms $F$ to $\tilde F \equiv e^{iS} F e^{-iS}$ such that $[\tilde F,F_0] = 0$ to the desired order in $V$. 
Expanding, we have  
\begin{align}
\label{eq:F_tilde}
\tilde F =& F_0 + i[S, F_0] + i F_0 V -\frac{1}{2} \{S^2, F_0 \}  \nonumber \\ 
&+ [F_0 V, S] + S F_0 S - \frac{1}{2} F_0 V^2 + \ldots .
\end{align}
We now write $S$ as a power series in $\lambda$, $S = \sum_n S_{(n)}$ with $S_{(n)} \in \mathcal O(\lambda^n)$. First, note that $S_{(0)} = 0$. This is because, to $0^{\text{th}}$ order in $V$, $F = F_0 + \mathcal O(V)$ already commutes with $F_0$. Expanding, the above expression becomes
\begin{align*}
\tilde F =& F_0 + (i[S_{(1)}, F_0] + i F_0 V) + ( i[S_{(2)}, F_0]-\frac{1}{2} \{S_{(1)}^2, F_0 \} \\ & +[F_0 V, S_{(1)}] + S_{(1)} F_0 S_{(1)} - \frac{1}{2} F_0 V^2) + \ldots,
\end{align*}
where we have grouped terms according to their order in $V$. Each of these grouped pieces self-consistently defines $S_{(n)}$. In particular, we require that each of them commute with $F_0$ to that order. In general, each self-consistent equation will be of the form 
\begin{equation}
\label{eq:Mn}
[ i [S_{(n)}, F_0] + M_{(n)} , F_0 ] = 0,
\end{equation}
where $M_n$ are the $n$th order terms obtained from expanding the expression for $\tilde F$: $M_{(1)} = i F_0 V$ and $M_{(2)} = -\frac{1}{2} \{S_{(1)}^2, F_0 \} + [F_0 V, S_{(1)}] + S_{(1)} F_0 S_{(1)} - \frac{1}{2} F_0 V^2$. In order to solve equations of the form of Eq. \ref{eq:Mn}, it is convenient to introduce the set of projectors $\{P_\alpha \}$  onto the eigenspaces of $F_0$, where $F_0 P_\alpha = \alpha P_\alpha$, $\alpha \in U(1)$ since $F$ is unitary, and $P_\alpha^2 = P_\alpha$. Then Eq. \ref{eq:Mn} can be readily solved as~\footnote{Note that the solution to Eq. \ref{eq:Mn} is only unique up to operators $\mathcal O$ satisfying $[[\mathcal O, F_0],F_0]=0$. Our solution is chosen to give blocks of zeros in $M_n$ corresponding to the eigenbasis of $F_0$.}

\begin{equation}
S_{(n)} = \sum_{\alpha \not= \beta} \frac{1}{i(\alpha - \beta)} P_\alpha M_{(n)} P_\beta.
\end{equation}

Having now described the framework for performing perturbation theory on a Floquet unitary,  the Floquet RSRG method proceeds as follows. First, identify the strongest piece $F_0$ in the Floquet evolution operator. This is similar to identifying the strongest bond $\Omega$ in the usual RSRG method; however, since quasi-energies take values on a circle,  they do not form a well-ordered set, and some care must be taken in identifying $\Omega$; we turn to this in the next section. For the moment, assume that such a  ``strong piece'' $F_0$ has been identified; then the rest of the chain is identified with $e^{iV}$. We then perform a Floquet SW transformation on $F$, truncating at second order in $V$. This generates a virtual coupling mediated by the strong piece $F_0$, giving a renormalized coupling between neighboring degrees of freedom. Iterating this procedure generates a flow of the couplings in the chain in much the same way as is in the usual RSRG method.

\section{\label{sec:Ising} Application: The Driven Ising Chain}

\begin{figure}
\includegraphics[width=\columnwidth]{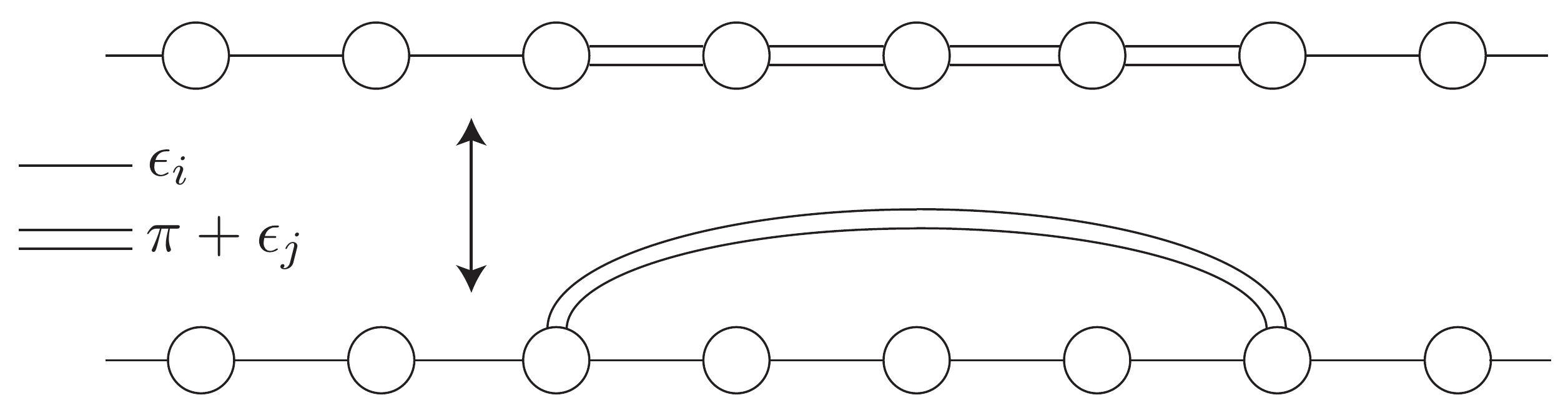}
\caption{\label{fig:factoring} (Top) A generic configuration of the disordered periodically driving Ising chain in the Majorana fermion representation. Circles represent Majorana fermion operators, single lines represent bilinear couplings $\epsilon_i$ with $\epsilon_i \in [-\pi/2,\pi/2]$, and double lines represent bilinear couplings $\pi + \epsilon_j$ with $\epsilon_j$ in the same range. The chain has domains with couplings near $0$ (``0-domains'') and with couplings near $\pi$ (``$\pi$-domains''). (Bottom) We can factor out the exact $\pi$ pulses to form one long-ranged exact $\pi$ pulse spanning the $\pi$-domain, with all couplings now in the range $[-\pi/2, \pi/2]$.}
\end{figure}

In order to demonstrate how Floquet RSRG works in practice, we now turn to a concrete application of the method outlined in the preceding section to a prototypical Floquet system: namely, the periodically driven Ising chain, defined by the sequence
\begin{equation}
H(t) \!=\! \begin{cases}
H_1 \!=\! \sum_i  \!h_i \sigma^x_i  + U^{xx}_i \sigma_i^x \sigma_{i+1}^x, & \!\!0 \leq t \leq T/2 \\
H_2 \!=\! \sum_i \!J_i \sigma^z_i \sigma^z_{i+1} +  U^{zz}_i \sigma_i^z \sigma_{i+2}^z, & \!\!T/2 \leq t \leq T,
\end{cases}
\end{equation}

where $\sigma_i^{x,z}$ are Pauli operators on site $i$, $J_i$ and $h_i$ are uncorrelated random variables, and $U$ corresponds to small interaction terms that respect the Ising $\Z_2$ symmetry generated by $G_{\rm Ising} = \prod_i \sigma_i^x$. For now, we will take $U^{xx}_i = U^{zz}_i =0$ and discuss the role of interactions in Section~\ref{sec:ints}. Applying a Jordan-Wigner transformation to rewrite the chain in terms of Majorana fermions~\cite{sachdev2007quantum}, we see that the Floquet evolution operator is
\begin{equation}
F = \exp( \frac{1}{2}\sum_{i} J_{2i} \gamma_{2i} \gamma_{2i+1}) \exp( \frac{1}{2} \sum_{i} J_{2i-1}\gamma_{2i-1} \gamma_{2i}),
\end{equation}
where we set $T=1$ for convenience. The odd Majorana bonds correspond to field terms for the spins ($h_i \leftrightarrow J_{2i-1}$) while the even Majorana bonds correspond to bond terms for the spins ($J_i \leftrightarrow J_{2i}$). The Majorana operators $\gamma_j$ obey $\gamma_j^\dagger = \gamma_j$, $\gamma_j^2 = 1$, and $\{\gamma_i, \gamma_j\} = 2\delta_{ij}$. We restrict the couplings to fall in a window of width $2\pi$, specifically the range $J_i \in [-\pi/2,3\pi/2)$ which is symmetric about $0$ and $\pi$. All couplings may be brought into this range by noting that $e^{(J_{ij} + 2 \pi n) \gamma_i \gamma_j/2} \propto e^{J_{ij} \gamma_i \gamma_j/2}$ for integer $n$, which will share the same eigenstates and hence the same phase structure.

This model hosts four phases~\cite{khemani_prl_2016}. Two are connected to static counterparts in the $T\to 0$ limit: (1) a trivial paramagnet, which is short-range correlated and does not exhibit spontaneous symmetry breaking (SSB); and (2) a spin glass phase which spontaneously breaks the Ising symmetry and is long-range correlated in time, or equivalently hosts a localized edge Majorana fermion mode at zero quasi-energy. Two phases are unique to the driven setting: (3) a ``$\pi$-spin glass'', which is long-range correlated and spontaneously breaks both the Ising symmetry and time translation symmetry, or equivalently hosts an Majorana edge mode at $\pi$ quasi-energy, and is often referred to as a ``time crystal''; and (4) a ``$0\pi$-paramagnet'', which has short-range bulk correlations, does not break the Ising symmetry, but does break time translation symmetry, and equivalently hosts edge Majorana modes at both 0 and $\pi$ quasi-energy. 

A generic configuration of the chain will have some couplings closer to $0$ and others closer to $\pi$. This leads to two types of domain: domains of couplings nearer to 0 (``$0$ domains''), and domains of couplings nearer to $\pi$ (``$\pi$-domains''). We will assume that in these domains, the couplings are either very close to $0$, or very close to $\pi$: even if this is not initially the case microscopically, we will see that this becomes true self-consistently after running the RG, {\it i.e.}, this is a property of the RG fixed points we are after. Note that Majorana fermions obey a simple evolution equation: $e^{\theta \gamma_i \gamma_j} = \cos\theta + \gamma_i \gamma_j \sin \theta$ for $i\not=j$. Therefore, within a $\pi$-domain we can factor out all of the $\pi$ pulses to simply extract one long-ranged pulse at strength exactly $\pi$: $(\gamma_0 \gamma_1 )( \gamma_1 \gamma_2) \ldots (\gamma_{L-1}\gamma_L ) = \gamma_0 \gamma_L = e^{\frac{\pi}{2} \gamma_0 \gamma_L}$.

We perform this factoring across the entire chain as a first step, as diagrammed in Figure~\ref{fig:factoring}. This is one of the most important steps in our procedure, as what remains are small couplings (controlling perturbation theory) that we will now show how to decimate, along with large non-local couplings that significantly modify the decimation rules from those of similar static Hamiltonians. From here on we use the notation $J_i$ to denote a Majorana coupling in the range $[-\pi/2,\pi/2)$, assuming all larger couplings have been factored out.

\subsection{Decimating inside a 0-domain or $\pi$-domain}

\begin{figure}
\includegraphics[width=\columnwidth]{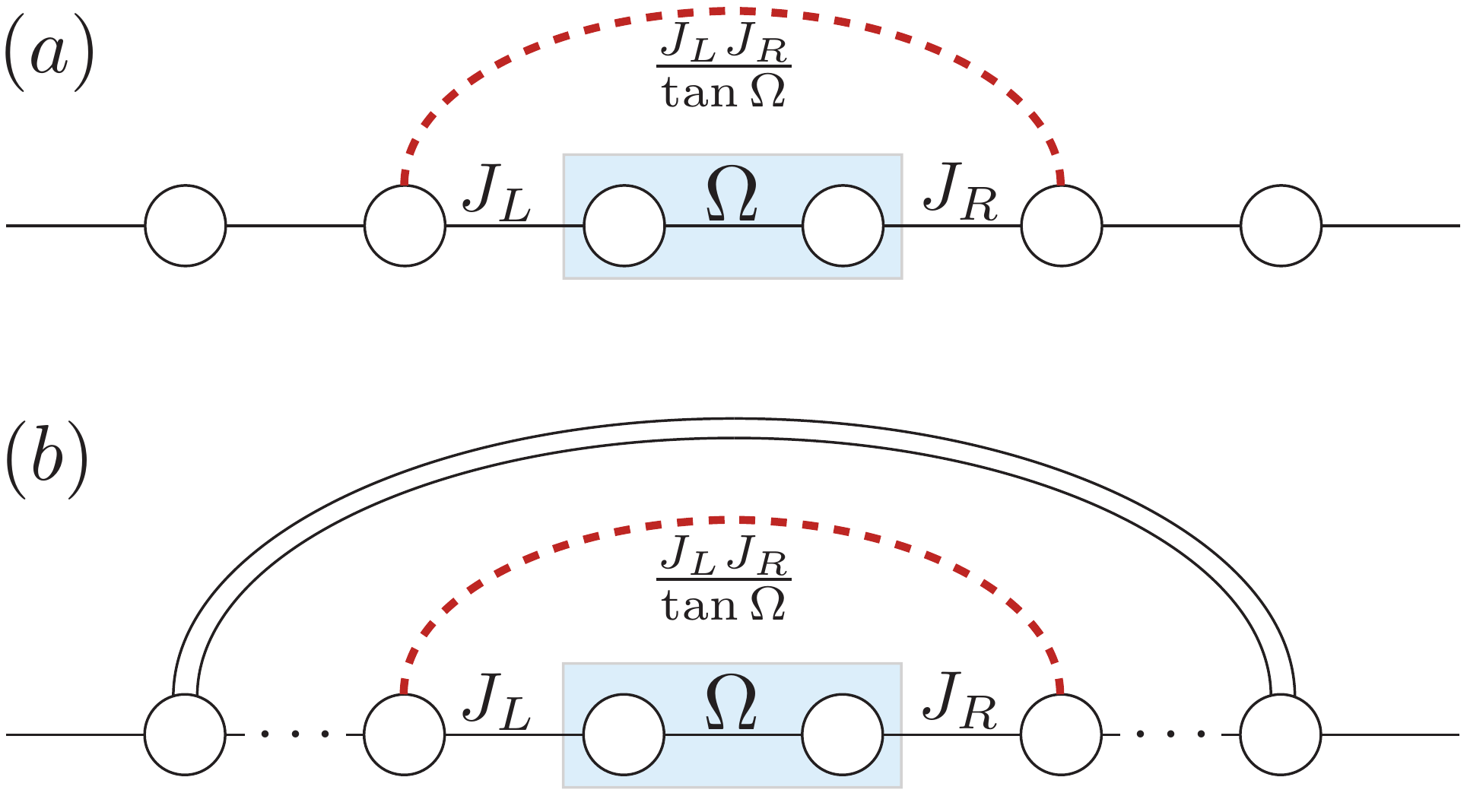}
\caption{\label{fig:domain_decimation} Decimation inside a 0-domain or a $\pi$-domain follows the same rule. (a) Within a 0-domain, a strong bond $\Omega$ is renormalized via the rule $\tilde J = J_L J_R / \tan \Omega$. (b) Within a $\pi$-domain, the long-ranged $\pi$ pulse does not affect the unitary rotation, giving the same virtual coupling. We can view decimation within a $\pi$-domain as decimation towards a $\pi$-bond by replacing the long-ranged $\pi$-pulse with a product of short-ranged ones.}
\end{figure}

As a warm-up exercise, consider decimation deep within a 0-domain or a $\pi$-domain. Assume that the strongest coupling is $J_0 \equiv \Omega \ll 1$, and that the neighboring couplings satisfy $J_L, J_R \ll \Omega \ll 1$, and are not on domain walls. This is actually the generic case deep in {\it both} a 0-domain and a $\pi$-domain; see Figure~\ref{fig:domain_decimation} for a diagram. We then have 
\begin{align}
F_0 &= e^{\Omega \gamma_0 \gamma_1} = \cos \Omega + \gamma_0 \gamma_1 \sin \Omega, \nonumber \\
V &= -i J_L \gamma_{-1} \gamma_0 - i J_R \gamma_0 \gamma_1.
\end{align}
All other terms commute with $F_0$, so we drop them. $F_0$ has two eigenstate projectors: $P_{\pm} = \frac{1}{2}(1 \pm i \gamma_0 \gamma_1)$ which project onto the $i \gamma_0 \gamma_1 = \pm 1$ subspaces, with eigenvalues $\lambda_\pm = e^{\mp i \Omega}$. Solving Eq.~\ref{eq:Mn} for $n=1,2$, we find
\begin{align}
S_{(1)} &= \frac{1}{2} i [ \gamma_{-1} \gamma_1 J_L \cot \Omega + \gamma_{-1} \gamma_0 J_L \\
&\qquad + J_R (\gamma_1 \gamma_2 - \gamma_0 \gamma_2 \cot \Omega) ], \nonumber \\
S_{(2)} &= 0.
\end{align}
Computing $P_+ \tilde F P_+$ and setting $i \gamma_0 \gamma_1 = 1$, 
\begin{align}
&P_+ \tilde F P_+ = \frac{1}{2} \sin\Omega (\cot \Omega - i) \nonumber \\ 
&\times[ -i(J_L^2 + J_R^2 ) \cot \Omega + 2 \gamma_{-1} \gamma_2 J_L J_R \cot\Omega + 2 ].
\end{align}
Now note that this is of form $P_+ \tilde F P_+ = e^{i \theta} e^{ \tilde J \gamma_{-1} \gamma_2} = e^{i \theta} (\cos \tilde J + \gamma_{-1} \gamma_2 \sin \tilde J ) = A + B \gamma_{-1} \gamma_2$. Thus, $\tan \tilde J = B / A$. Looking above to identify $A$ and $B$ and dividing, we find
\begin{equation}
\label{eq:deep_rule}
\tan \tilde J = \frac{J_L J_R}{\tan \Omega - \frac{i}{2} (J_L^2 + J_R^2) } \approx \frac{J_L J_R}{\tan \Omega}.
\end{equation}

That is, the renormalized coupling is related to the {\it tangent} of the strong bond $\Omega$; this reduces to the well-known static rule $\tilde J = J_L J_R / \Omega$ in the limit $\Omega \to 0$. We can similarly project onto the $P_-$ subspace to access other branches of the many-body spectrum, finding the same rule of $\tan \tilde J = J_L J_R / \tan \Omega$ but a different overall quasi-energy shift (Eq.~\ref{eq:quasi-energy}). Though Ref.~\onlinecite{1742-5468-2017-7-073301} obtains a superficially similar formula, we note that in contrast to Eq.(45) of Ref.~\onlinecite{1742-5468-2017-7-073301}, Eq.~\ref{eq:deep_rule} is valid when all nearby bonds are near $\pi$ {\it as well as} near $0$, due to the factoring of the $\pi$ pulses.

One crucial difference with the static case is that the ordering of terms is now important; one might wonder if the proposed renormalization procedure even gives back a self-similar evolution operator for this model. Indeed it does, as can be seen by examining $F$. Taking $\Omega$ in the even sublattice for the sake of argument, let us decimate $\Omega \gamma_0 \gamma_1$:
\begin{align}
&F = \ldots e^{J_{-2} \gamma_{-2} \gamma_{-1} + J_{2} \gamma_2 \gamma_3} e^{\Omega \gamma_0 \gamma_1} \nonumber \\ 
 & \qquad \times e^{J_{-1} \gamma_{-1} \gamma_0 + J_{1} \gamma_1 \gamma_2} e^{J_{-3} \gamma_{-3} \gamma_{2} + J_{3} \gamma_3 \gamma_4} \ldots \nonumber \\
&\xrightarrow{\text{RG}} \ldots e^{J_{-2} \gamma_{-2} \gamma_{-1} + J_{2} \gamma_2 \gamma_3} e^{\tilde J \gamma_{-1} \gamma_2 + i \theta} e^{J_{-3} \gamma_{-3} \gamma_{2} + J_{3} \gamma_3 \gamma_4} \ldots \nonumber \\
&= \ldots e^{i \theta} e^{J_2 \gamma_2 \gamma_3 + J_{-2} \gamma_{-2} \gamma_{-1}} e^{J_{-3} \gamma_{-3}\gamma_{-2} + \tilde J \gamma_{-1}\gamma_2 + J_3 \gamma_3 \gamma_4} \ldots ,
\end{align}
where in the first step we factored commuting pieces, and $e^{i\theta}$ is the overall quasi-energy shift. Odd sublattice decimations proceed similarly. Therefore, we see that the decimation has produced a self-similar Floquet operator, and decimating a bond in the even (odd) sublattice produces a renormalized bond in the odd (even) sublattice, as in the static case. 

We can also determine how much the quasi-energy has shifted by isolating $e^{i\theta}$. Since we expect a renormalized Floquet operator of form $e^{i \theta}e^{\phi \tilde{\mathcal O} }$, with $\tilde{\mathcal O} $ some operator, we recover $e^{i\theta}$ by simply taking $\tilde{\mathcal O} \to 0$, or equivalently $\gamma_i \to 0$ for all $i$. Thus,
\begin{equation} \label{eq:quasi-energy}
e^{i \theta} = \frac{(J_L^2 + J_R^2) \cos \Omega + 2i c\sin\Omega}{e^{i 2 c \Omega} - 1},
\end{equation}
where $c = \pm 1$ picks the $P_+$ or $P_-$ branch, respectively. In the limit $\Omega \ll 1$, we can expand this expression and check that it reproduces the energy shift in the static case~\cite{PhysRevX.4.011052}: $\theta \approx c ( \frac{J_L^2 + J_R^2}{2 \Omega} + \Omega ) + \mathcal O (\Omega^2)$, as it should.  \\

Thus, for decimations that do not encounter a domain wall, we obtain a straightforward generalization of the usual RSRG decimation rules for the static Ising chain. Note that, if we are within a $\pi$-domain, we can replace the long-range $\pi$-pulse by a product of short range $\pi$-pulses at any time by just reversing the factoring argument above section A (see Figure~\ref{fig:factoring}). This tells us that after we decimate within a $\pi$-region, we are actually decimating the bond towards $\pi$. Therefore, the domain type is maintained under renormalization.

\subsection{Decimating near a domain wall}

\begin{figure}
\includegraphics[width=\columnwidth]{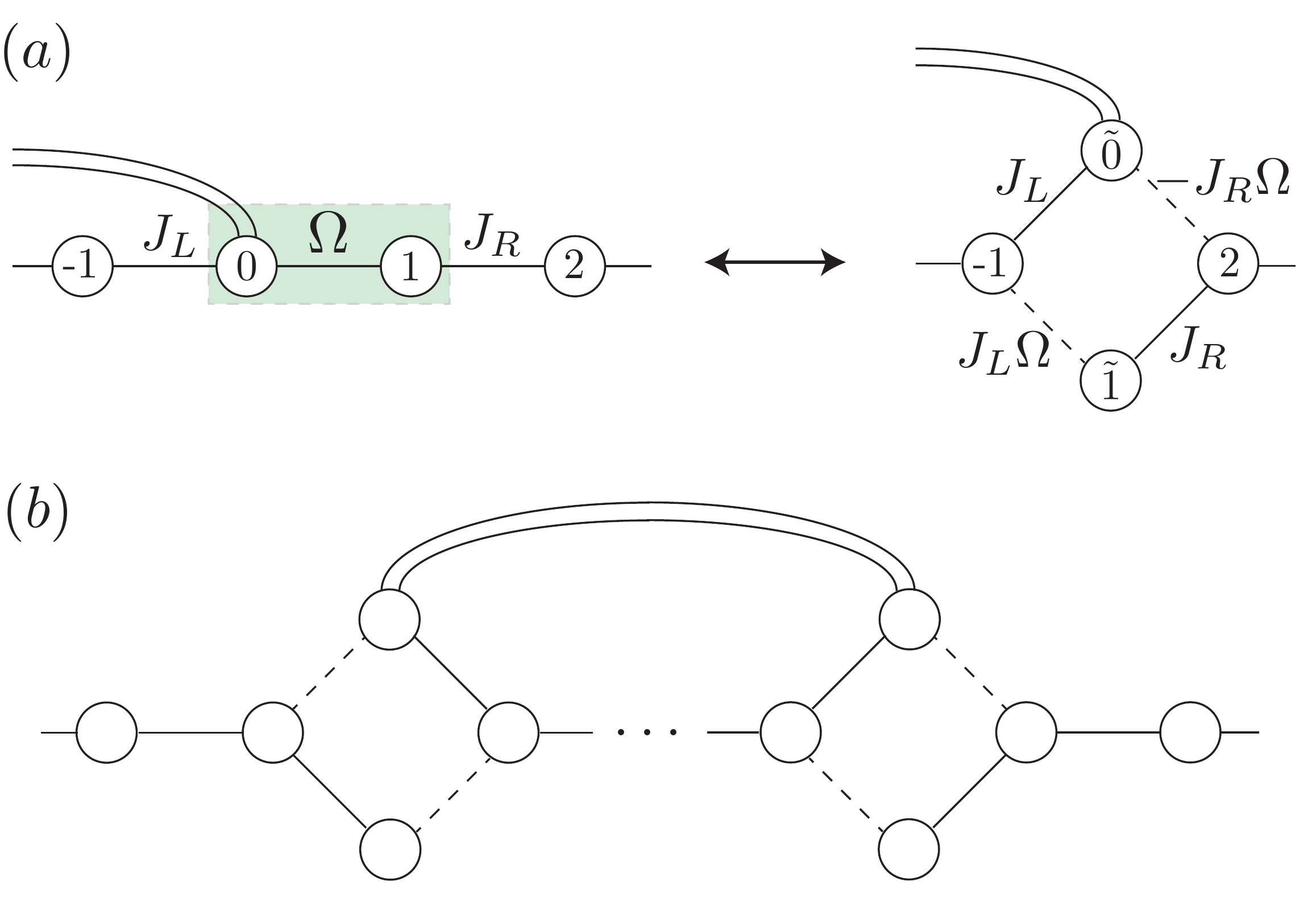} \\
\caption{\label{fig:dw} (a) An exact transformation of each domain wall reveals that we can view the $\pi$ and 0 puddles as being connected by 2nd order terms (see below for details). We will never decimate these because $J_R \Omega < J_R$ and $J_L \Omega < J_L$, so at every step $J_L$ or $J_R$ dominate. (b) The chain after the exact transformations of factoring out $\pi$-pulses and rotating the Majoranas at every domain wall. Approximating $\cos\Omega \approx 1, \sin \Omega \approx \Omega$ shows that the puddles are decoupled at leading order.}
\end{figure}

Now let us consider the interesting case where we want to decimate a bond on the domain wall between a 0-domain and a $\pi$-domain. Let us imagine that the strongest bond in the chain lies at the domain wall between a 0-domain and a $\pi$-domain. That is, let the bonds $J_j$ with $-L<j<0$ be near $\pi/2$, and other bonds near 0. The Floquet operator is
\begin{equation}
F = \ldots e^{\Omega \gamma_0 \gamma_1} \gamma_{-L} \gamma_0 e^{J_L \gamma_{-1}\gamma_0 + J_R \gamma_1 \gamma_2} \ldots 
\end{equation}
where we have pulled out the $\pi$-pulses within the $\pi$-domain. Focusing on the domain wall at site $0$, note that the Floquet operator in this vicinity is $e^{\Omega \gamma_0 \gamma_1} \gamma_{-L} \gamma_0 e^{J_L \gamma_{-1}\gamma_0}$. Let us assume that $\Omega > J_L$; then naively we may wish to identify the ``strong'' piece as $F_0 = e^{\Omega \gamma_0 \gamma_1} \gamma_{-L} \gamma_0$. However, the eigenvalues of this operator actually {\it do not depend} on $\Omega$, so perturbation theory in $\Omega^{-1}$ is not well-controlled. To see this, note that the middle piece can be factored as $ \exp(\Omega \gamma_0 \gamma_1) \gamma_{-L} \gamma_0 =\gamma_{-L} (\gamma_0 \cos\Omega - \gamma_1 \sin \Omega)$. We can define rotated Majorana operators
\begin{equation}
\begin{pmatrix} \tilde \gamma_0 \\
\tilde \gamma_1 \end{pmatrix} = \begin{pmatrix} \cos\Omega & -\sin \Omega \\
\sin \Omega & \cos \Omega \end{pmatrix} \begin{pmatrix} \gamma_0 \\ 
\gamma_1 \end{pmatrix} \approx \begin{pmatrix} 1 & -\Omega \\
\Omega & 1 \end{pmatrix}  \begin{pmatrix} \gamma_0 \\ 
\gamma_1 \end{pmatrix} 
\end{equation}
since $\Omega \ll 1$ (we factored out all $\pi$-pulses). One can verify that this gives a new set of Majorana fermions: $\tilde \gamma_{0,1}^\dagger = \tilde \gamma_{0,1}$, $\tilde \gamma_{0,1}^2 = 1$, $\{\gamma_i, \tilde \gamma_{0,1}\}_{i\not=0,1} = 0$, and $\{\tilde \gamma_0, \tilde \gamma_1\} = 0$. Rewriting in terms of these variables,
\begin{align}
F = \ldots \gamma_{-L} \tilde \gamma_0 \exp(&J_L \gamma_{-1} \tilde \gamma_0 + J_L \Omega \gamma_{-1} \tilde \gamma_1 \nonumber \\
 - &J_R \Omega \tilde \gamma_0 \gamma_2 + J_R \tilde \gamma_1 \gamma_2) \ldots,
\end{align}
where we have eliminated all couplings of strength $\Omega$ in favor of weaker terms.
This transformed picture is diagrammed in Figure~\ref{fig:dw}(a). We note that we could have just as easily performed a similar rotation based around $J_L \gamma_{-1} \gamma_0$, but find it more convenient to always choose to rotate the bond that is closer to 0 than to $\pi$. In Ref.~\onlinecite{potter_classification_2016}, a similar rotation argument was given to demonstrate that bilinear couplings between zero- and $\pi$-quasienergy Majorana fermions cannot change their quasienergy.

\begin{figure}
\includegraphics[width=\columnwidth]{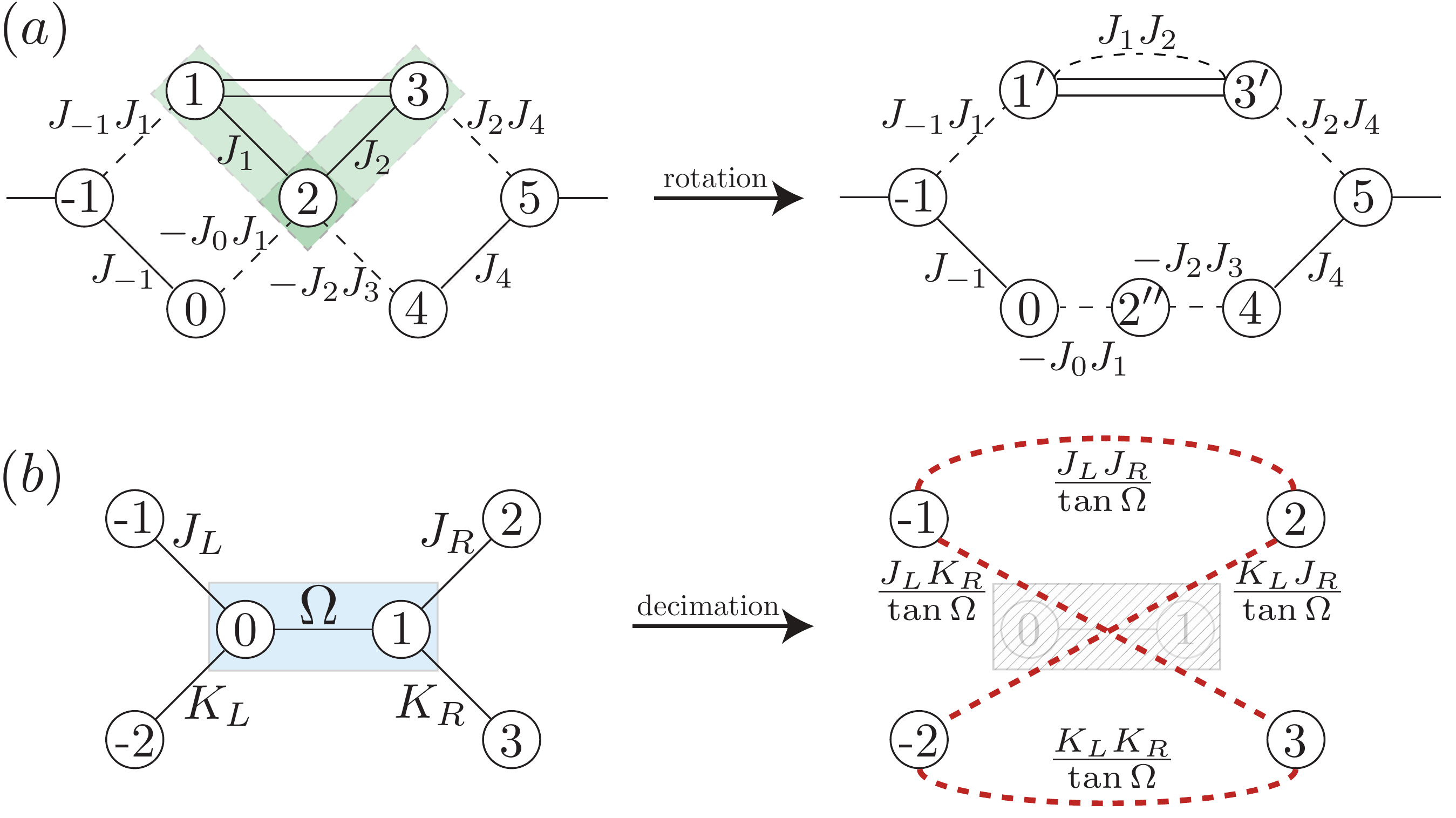}
\caption{\label{fig:pathway} (a) An even-length domain, after it has been decimated down to only two bonds, can be rotated into this two-chain form without decimating. (b) When decimating the central bond in the setup above, every three-step pathway through the decimated bond gets renormalized according to $M_L M_R/\tan \Omega \approx M_L M_R/\Omega$, where $M_{L,R}$ are the bonds to the left and right respectively. This type of decimation occurs with odd-length domains.}
\end{figure}

\begin{figure*}[ht!]
\includegraphics[width=2.\columnwidth]{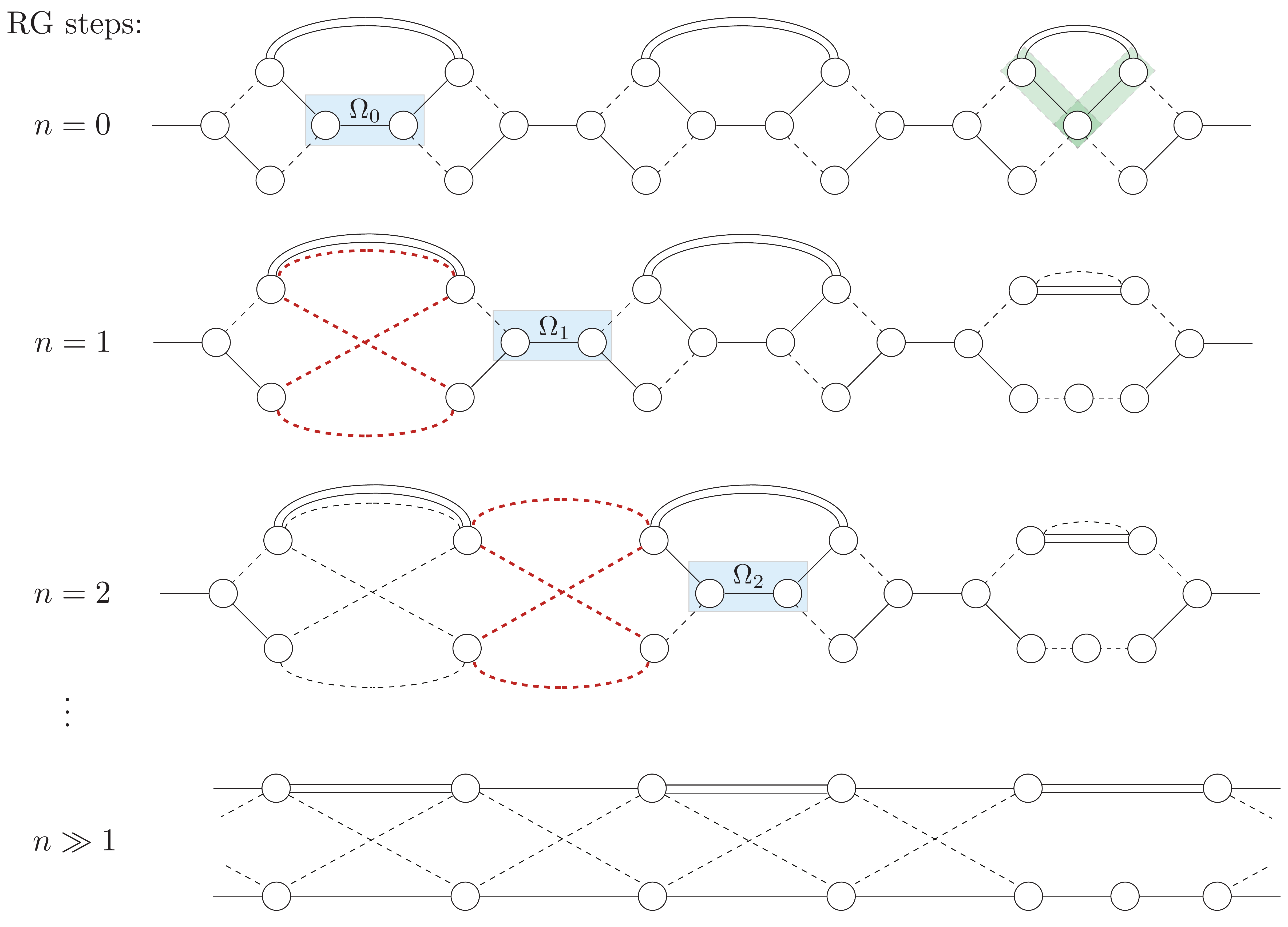}
\caption{\label{fig:two_chains} After the 0 and $\pi$-domains have been decimated, we are left with a configuration of similar form to the chain in (1). As we decimate the central couplings of each puddle, we renormalize to the chain in step (2), and so on. This leads to two linked chains.}
\end{figure*}

The above argument applies equally well to the bond $J_L \gamma_{-1}\gamma_0$, so we find that neither bond touching the domain wall Majorana should be (naively) decimated. This has a physical interpretation: decimating either bond is akin to decimating the topological edge mode that lies at this interface between a 0 region and a $\pi$ region, so to lowest order is at exactly $\pi$ quasi-energy independent of the surrounding coupling strengths. Since these edge modes are at 0 or $\pi$ quasi-energy, perturbation theory is not well-controlled when attempting to decimate them. Since, as previously argued~\cite{Berdanier201805796}, these topological edge modes control the criticality, it would not be sensible to decimate them outright, and the surrounding bonds will only be decimated at higher order, later in the RG.

\subsection{Decimations between the $0$- and $\pi$-chains}

Let us run the above RG until all of the 0-domain and $\pi$-domains have been decimated and we are left with many domain walls. Note first that we must distinguish between domains that began with an even number of bonds versus those that began with an odd number. Since a decimation removes two Majoranas, after many decimations an odd-length puddle will be left with three bonds, while an even-length puddle will be left with only two. We find it convenient to first perform a rotation on the two-bond puddles, described below. 

Let us consider a two-bond $\pi$-domain, namely $J_j \gamma_j \gamma_{j+1}$ which has $J_j \approx \pi$ for $j=0,1$ and $J_j$ near 0 otherwise. We first rotate this domain's domain walls as above, constructing $\tilde \gamma_{0,1} = \gamma_{0,1} \pm J_0 \gamma_{1,0}$ and $\gamma_{1,2}' = \gamma_{1,\tilde 2} \mp J_1 \gamma_{\tilde 2,1}$. This gives a domain structure as diagrammed in Fig.~\ref{fig:dw}b and the left part of Fig.~\ref{fig:pathway}(a), dropping tildes and primes for clarity. Then let us define two consecutive rotations, $\gamma_{0}' = \gamma_{0} + J_0 \gamma_{1}$, $\gamma_{1}' = \gamma_{1} - J_0 \gamma_{0}$, followed by $\gamma_{1}'' = \gamma_{1}' - J_1 \gamma_{2}$, $\gamma_{2}'' = \gamma_{2} + J_1 \gamma_{1}'$. Dropping third order terms, this leads to the following bonds: $J_{-1}J_1 i \gamma_{-1} \gamma_1$, $(\pi/2 + J_1 J_2) i \gamma_1' \gamma_3'$, $J_{-1} i \gamma_{-1} \gamma_0$, $-J_0 J_1 i \gamma_0 \gamma_2''$, $-J_2 J_3 \gamma_2'' \gamma_4$, $J_4 i \gamma_4 \gamma_5$, and $J_2 J_4 i \gamma_3' \gamma_5$, as diagrammed in the right half of Fig.~\ref{fig:pathway}(a). Therefore, with a two-bond puddle it is sensible to rotate the bonds so that the puddle is of two-chain form and each bond can be straightforwardly decimated. 

Now consider an odd-length puddle. As we run the RG, we avoid decimating the domain-wall bonds, so the puddle size shrinks until there are just three bonds remaining. These 3-bond puddles will be left with a configuration similar to Figure~\ref{fig:two_chains} at RG step 1. We now want to decimate the central bond of one of the puddles. Let us calculate this rule, diagrammed in Figure~\ref{fig:pathway}(b). The Floquet operator is of the form
\begin{equation}
F = \ldots e^{\Omega \gamma_0 \gamma_1} e^{J_L \gamma_{-1} \gamma_0 + K_L \gamma_{-2} \gamma_0+J_R \gamma_1 \gamma_2 + K_R \gamma_1 \gamma_3} \!\ldots\!\!
\end{equation}
so we identify
\begin{align}
F_0 &= e^{\Omega \gamma_0 \gamma_1} \\
V &= -i J_L \gamma_{-1} \gamma_0 - i K_L \gamma_{-2} \gamma_0 - i J_R \gamma_1 \gamma_2 - i K_R \gamma_1 \gamma_3\nonumber.
\end{align}

Applying the machinery from above, we arrive at a renormalized Floquet operator where every 3-step pathway through the central bond is renormalized as $M_L M_R / \tan\Omega$, where $M_{L,R}$ are the bonds on the left and right of $\Omega$, respectively. That is,
\begin{align}
\tilde F = \ldots \exp \Bigg( &\frac{J_L J_R}{\tan\Omega} \gamma_{-1} \gamma_2 + \frac{J_L K_R}{\tan\Omega} \gamma_{-1} \gamma_3 \nonumber \\
+ &\frac{K_L J_R}{\tan\Omega} \gamma_{-2} \gamma_2 + \frac{K_L K_R}{\tan\Omega} \gamma_{-2} \gamma_3 \Bigg)\ldots
\end{align}
with quasi-energy shift
\begin{equation}
e^{i \theta} = \frac{\cos\Omega (J_L^2 + J_R^2 + K_L^2 + K_R^2) + 2 i c \sin \Omega}{e^{i 2 c \Omega} - 1},
\end{equation}
where again $c=\pm 1$ specifies the branch choice. Applying this decimation rule repeatedly will lead to two connected chains, as shown in Figure~\ref{fig:two_chains}. 

Note that the bottom chain has all couplings near 0 and hence is near 0 quasi-energy, while the top chain has every other coupling near $\pi$, and hence is at $\pi$-quasi-energy~\footnote{This can be verified by comparison with the microscopic phase diagram of the model, as shown in Ref.~\onlinecite{Berdanier201805796}}. We might worry that the couplings connecting the chains will need to be decimated. However, they cannot be decimated for the same reason that we could not directly decimate a bare domain-wall coupling earlier: each of these links has a coupling near $\pi$ to one side and a coupling near 0 to the other. Thus, at leading order the energetics are independent of these couplings, and we can perform a Majorana rotation as above to push them to higher order. Once we do this, we find that the higher order couplings actually {\it still} connect a 0 coupling on one side to a $\pi$ coupling on the other, and can be rotated yet again to push these couplings to an even higher order. Ultimately, if we continue to rotate ad infinitum, we will decouple the chains without doing any decimations. This leaves us with two effectively decoupled chains, one at 0 and one at $\pi$ quasi-energy. 

{
We can write flow equations, using the rules above, for the four distributions in the problem: the distributions of bonds and fields near 0 and $\pi$. Define logarithmic variables $\Gamma\equiv \log(\Omega_I / \Omega)$, which sets the overall RG scale for some arbitrary initial scale $\Omega_I$; $\zeta^{0,\pi} \equiv \log(\Omega/J^{0,\pi})$, where $J$ is from the even sublattice in terms of Majoranas; and $\beta^{0,\pi} \equiv \log(\Omega/h^{0,\pi})$, where $h$ is from the odd sublattice. These last four logarithmic variables have four associated coupling distributions: $P_{\beta,\zeta}^{0,\pi}$. Then the above decimation rules translate to sum rules in terms of logarithmic variables {at strong disorder}: $\tilde \beta^\theta = \zeta^\theta_L + \zeta^\theta_R$, $\tilde \zeta^\theta = \beta^\theta_L + \beta^\theta_R$, with $\theta = 0, \pi$ indexing the two chains. These rules give rise to two coupled RG flow equations for each chain~\cite{PhysRevB.51.6411}
\begin{align}
\pd{P^\theta_\zeta}{\Gamma}(\zeta^\theta) &= \pd{P^\theta_\zeta}{\zeta^\theta}(\zeta^\theta) + P_\beta^\theta(0) \int_0^\infty d\bar\zeta^\theta P_\zeta^\theta(\bar \zeta^\theta) P_\zeta^\theta(\zeta^\theta - \bar \zeta^\theta) \nonumber \\
&\quad + P_\zeta^\theta(\zeta^\theta) [ P_\zeta^\theta(0) - P_\beta^\theta(0) ], \nonumber \\
\pd{P^\theta_\beta}{\Gamma}(\beta^\theta) &= \pd{P^\theta_\beta}{\beta^\theta}(\beta^\theta) + P_\zeta^\theta(0) \int_0^\infty d\bar\beta^\theta P_\beta^\theta(\bar \beta^\theta) P_\beta^\theta(\beta^\theta - \bar \beta^\theta) \nonumber \\
&\quad + P_\beta^\theta(\beta^\theta) [ P_\beta^\theta(0) - P_\zeta^\theta(0) ],
\end{align}
with $\theta = 0,\pi$. These equations give rise to the usual infinite-randomness fixed-point distributions $\tilde P_{\beta}^\theta(\beta^\theta) = \frac{1}{\Gamma} e^{-\beta^\theta / \Gamma}$, $\tilde P_{\zeta}^\theta(\zeta^\theta) = \frac{1}{\Gamma} e^{-\zeta^\theta / \Gamma}$ {at criticality}, showing flow to two IRFPs at 0 and $\pi$ quasienergy.}

\begin{figure}
\includegraphics[width=0.8\columnwidth]{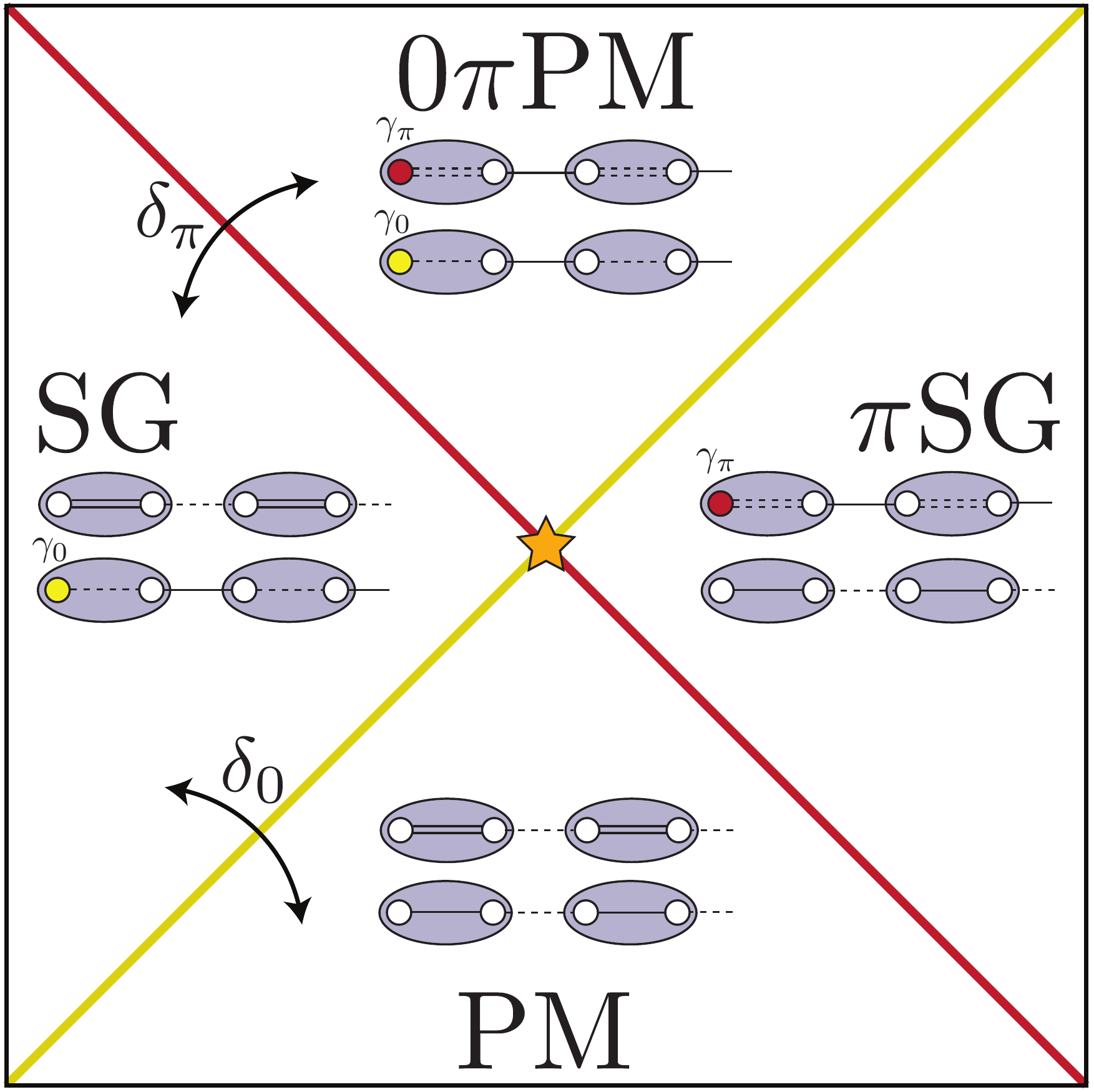}
\caption{\label{fig:phases} Cartoon of the four possible phases and their transitions as deduced from the strong-disorder RG flow to two decoupled Majorana chains near 0 and $\pi$ quasienergy. The phases are controlled by the dimerization of these chains $\delta_{0,\pi}$. The PM has both chains trivially dimerized; the SG has the 0-chain topologically dimerized and the $\pi$-chain trivially dimerized; the $\pi$SG has the 0-chain trivially dimerized and the $\pi$-chain topologically dimerized; and the $\pi$SG has both chains topologically dimerized (insets). Both chains can be independently tuned to Ising-class criticality (yellow and red lines), resulting in a twice-Ising multicritical point (orange star).
}
\end{figure}

Now that we have flowed to two decoupled Ising chains, the four possible phases of the model become clear: each chain can be independently dimerized, with a dimerization parameter $\delta_{0,\pi}$ controlling the phases. If the 0 ($\pi$)-chain is dimerized in the trivial pattern $\delta_{0(\pi)} < 0$, there is no edge mode, while if the 0 ($\pi$)-chain is dimerized in the topological pattern $\delta_{0(\pi)} > 0$, there is an edge Majorana mode at 0 ($\pi$) quasienergy. The four phases are thus identified by the sign of the dimerization patterns at $0,\pi$; accordingly we may label the phases by $(\delta_0,\delta_\pi)$ where $\delta_{0\pi} = +,-$. With this convention,  we identify the PM as both chains trivially dimerized $(-,-)$; the SG as the 0 chain topologically dimerized and the $\pi$-chain trivially dimerized, $(+,-)$; the $\pi$SG as the 0-chain trivially dimerized and the $\pi$-chain topologically dimerized $(-,+)$; and the $0\pi$PM as both chains topologically dimerized $(+,+)$. The critical lines are then set by $\delta_{0,\pi} = 0$ with both vanishing at the multicritical point of the model (see Fig.~\ref{fig:phases}). One can tune between these phases microscopically by adjusting the ratio of couplings near 0 and $\pi$.~\cite{Berdanier201805796}

This two-chain structure also elucidates the model's (multi)criticality. Each chain can be tuned to criticality independently, and each will be in the Ising universality class~\cite{PhysRevB.51.6411,PhysRevLett.69.534}. In particular, this implies that the coupling fixed-point distribution will be in the form of a stretched exponential $\sim e^{-1/w}$, where $w$ sets the disorder strength and $w\to \infty$ under renormalization when the system is at criticality ($\delta = 0$). This infinite randomness fixed point has dynamical exponent $z = 1 / 2 \abs{\delta} \to \infty$, characteristic of slow, glassy scaling. As one tunes slightly away from criticality, one finds that the correlation length diverges as $\xi \sim \abs{\delta}^{-\nu}$, with $\nu = 2$ or $\nu = 1$ for average ($\avg{\mathcal O}$) or typical ($e^{\avg{\log\mathcal O}}$) quantities, respectively. Further results include the critical spin-spin correlation function scaling $\avg{ \expectation{\sigma_j^z \sigma_{j+x}^z}} \sim 1/x^{2-\phi}$, with $\phi = (1+\sqrt{5})/2$ the golden ratio, and the entanglement entropy scaling $\overline{S_\ell} = (\tilde c / 6) \log \ell$ with $\tilde c = (1/2) \ln 2$ with open boundary conditions for a cut from the boundary of length $\ell$.~\cite{PhysRevLett.93.260602} At the system's multicritical point, we find that the chain flows under renormalization to two chains, one at 0 and the other at $\pi$ quasi-energy, each in the random Ising universality class.

\section{Interactions \label{sec:ints}}

Now that we have addressed the free problem, let us return to adding small but finite interactions $U \ll 1$. As a reminder, we introduced interaction terms such that the first piece of the drive was $H_1 = \sum_i h_i \sigma_i^x + U^{xx}_i \sigma_i^x \sigma_{i+1}^x$ and the second piece was $H_2 = \sum_i J_i \sigma_i^z \sigma_{i+1}^z + U^{zz}_i \sigma_i^z \sigma_{i+2}^z$. In terms of Majoranas, these interactions are statistically~\footnote{Absent disorder, the self-duality would be exact, but here it is only true in a statistical sense, i.e. it is to be understood as a duality map of various observables that occurs when the different couplings are drawn from the same distribution.}  {self-dual}
 under the usual Ising bond-field duality that interchanges $\sigma^x_i \leftrightarrow \sigma^z_i \sigma^z_{i+1}$. Therefore bond-field duality exchanges the operator content of $H_1$ and $H_2$, while preserving coefficients. Upon performing a Jordan-Wigner transformation for these interaction terms, and redefining $J_{2i} \equiv h_i$, $J_{2i+1} \equiv J_i$, $U_{2i} \equiv U^{xx}_i$, $U_{2i+1} \equiv U^{zz}_i$, the full Floquet operator reads 
\begin{align}
F &= e^{\sum_j J_{2j} i \gamma_{2j} \gamma_{2j+1} - U_{2j} \gamma_{2j} \gamma_{2j+1} \gamma_{2j+2} \gamma_{2j+3} } \nonumber \\
&\times e^{\sum_j J_{2j+1} i \gamma_{2j+1} \gamma_{2j+2} - U_{2j+1} \gamma_{2j+1} \gamma_{2j+2} \gamma_{2j+3} \gamma_{2j+4} }.
\end{align}

Let us first discuss decimating deep in one of the 0-domain or $\pi$-domains, such that all bonds are near 0. Then let us say that the strongest bond is $J_3$, such that 
\begin{align}
F_0 &= e^{J_3 \gamma_3 \gamma_4}, \\
V &= i J_0 \gamma_0 \gamma_1 + i J_1 \gamma_1 \gamma_2 + i J_4 \gamma_4 \gamma_5 + i J_5 \gamma_5 \gamma_6 + i J_6 \gamma_6 \gamma_7 \nonumber \\
& - U_0 \gamma_{0123} - U_1 \gamma_{1234} - U_2 \gamma_{2345} - U_3 \gamma_{3456} - U_4 \gamma_{4567},
\end{align}
where $\gamma_{ijk\ldots l} \equiv \gamma_i \gamma_j \gamma_k \ldots \gamma_l$ for short. Performing the Floquet Schrieffer-Wolff transformation and projecting onto the subspace $i\gamma_3 \gamma_4 = c$ with $c=\pm 1$, we find, after some algebra, that the renormalized operator is 
\begin{align}
\tilde F &= \ldots e^{i \theta} e^{ \tilde J_0 \gamma_0 \gamma_1 + \tilde J_2 \gamma_2 \gamma_5 + \tilde J_6 \gamma_6 \gamma_7 +  \tilde U_0 i \gamma_{0125} + \tilde U_2 i \gamma_{2567} + \tilde W \gamma_{012567} } \nonumber \\ 
&\times e^{ \tilde J_1 \gamma_1 \gamma_2 + \tilde J_5 \gamma_5 \gamma_6 + \tilde U_1 i \gamma_{1256} } \ldots 
\end{align}
with the following decimation rules:
\begin{align}
&\tilde J_0 = J_0 + c \frac{J_2 U_0}{\tan J_3}, 
~~~ &\tilde J_1 &= J_1 + c U_1, \nonumber \\
&\tilde J_2 = \frac{J_2 J_4}{\tan J_3} + U_2, 
~~~ &\tilde J_5 &= J_5 + c U_3, \nonumber \\
&\tilde J_6 = J_6 + c \frac{J_4 U_4}{\tan J_3}, 
~~~ &\tilde U_0 &= \frac{J_4}{\tan J_3} U_0, \nonumber \\
&\tilde U_1 = 0, 
~~~ &\tilde U_2 &= \frac{J_2}{\tan J_3} U_4, \nonumber \\
&\tilde W = \frac{U_0 U_4}{\tan J_3},
\end{align}
and overall quasi-energy shift
\begin{align}
e^{i \theta} = &\frac{1}{2} e^{-i c J_3} [ 2- i c \frac{J_2^2 + J_4^2 + U_0^2 + U_4^2}{\tan J_3} \nonumber  \\
&  -J_0^2 - (J_1 + U_1)^2- (J_5 + U_3)^2 - J_6^2 - U_2^2 ].
\end{align}

First, note that in the limit $\tan J_3 \approx J_3$, we recover the decimation rules of the static case~\footnote{These static case rules are derived in the presence of only $\sigma^x_i \sigma^x_{i+1}$ interactions in Appendix B of Ref.~\onlinecite{PhysRevX.4.011052}, though we note that the published version contains several sign errors.}, as we must. A six-fermion term $-i \tilde W \gamma_{012567}$ is generated in the Schrieffer-Wolff transformation that is second order in the interaction strength, which is already taken to be weak. The generation of fermion strings from RSRG methods has already been explored in the static case~\cite{PhysRevLett.112.217204}, where it was found that as the RG progresses, fermion strings of length $2m$ ($\sigma^x$ strings of length $m$) were generated at order $J (J/h)^{m-2}$. Since these coefficients are exponentially decaying with $m$, they can be safely discarded, though one can keep track of them if one wishes. We find that they exponentially decay here as well, so we neglect these strings and keep only fermion bilinears and four-fermion interaction terms. 

Now, consider the role of interaction terms at the domain wall. Let us rotate $\gamma_{2,3}$ to form $\tilde \gamma_2 = \gamma_2 + J_{2} \gamma_3$, $\tilde \gamma_3 = \gamma_3 - J_{2} \gamma_2$. Then a four-fermion interaction term $U_0 \gamma_{0123}$ will rotate to $U_0 \gamma_0 \gamma_1 (\tilde \gamma_2 - J_2 \tilde \gamma_3)(\tilde \gamma_3 + J_2 \tilde \gamma_2) = U_0 \gamma_0 \gamma_1 ( \tilde \gamma_2 \tilde \gamma_3 - J_2 \tilde \gamma_3^2 + J_2 \tilde \gamma_2^2 - J_2^2 \tilde \gamma_3 \tilde \gamma_2 ) = U_0 \gamma_0 \gamma_1 \tilde \gamma_2 \tilde \gamma_3 ( 1 + J_2^2)$. Therefore, the effect of the rotation is simply to renormalize the interaction to second order, as $U_0 (1+J_2^2) \gamma_{0 1 \tilde{2} \tilde{3}} \approx U_0 \gamma_{0 1 \tilde{2} \tilde{3}}$. Thus if the interaction term involves both of the rotated Majorana operators, the interaction is essentially unchanged. If it only involves one, on the other hand, we generate a new longer-ranged interaction term at one higher order: $U_{-1} \gamma_{-1012} = U_{-1}  \gamma_{-101\tilde 2} + U_{-1} J_2 \gamma_{-101\tilde 3}$. Finally, if an interaction term spans two separate rotations we obtain the following: $U_3 \gamma_{3456} = U_3 (\tilde \gamma_3 + J_2 \tilde \gamma_2)\gamma_4 \gamma_5 (\tilde \gamma_6 - J_2 \tilde \gamma_7) = U_3 \gamma_{\tilde{3} 45 \tilde{6}} + U_3 J_2 \gamma_{\tilde{2} 45 \tilde{6}} - U_3 J_6 \gamma_{\tilde{3} 45 \tilde{7}} - J_{2} J_{6} \gamma_{\tilde{2} 45 \tilde{7}}$. That is, we couple the Majoranas involved in both domain walls at leading order $\mathcal O(U)$ for the two nearest, one higher order $\mathcal O(UJ)$ for the next nearest, and one higher order still $\mathcal O(UJ^2)$ for the farthest away.

Having seen that no fundamentally new interactions are generated by this rotation, let us consider a bilinear decimation in the most general case of an odd-length domain, similar to that shown in Fig.~\ref{fig:pathway}(b). Then the Floquet operator will take the form 
\begin{align*}
  F =& \exp( J_0 \gamma_{01} + J_2 \gamma_{23} + \Omega \gamma_{45} + J_6 \gamma_{67} + J_8 \gamma_{89} + \ldots \\
  &~~~-i U_0 \gamma_{0123} -i U_2 \gamma_{2345} -i U_4 \gamma_{4567} -i U_6 \gamma_{6789} + \ldots ) \\ & \times \exp(J_1 \gamma_{12} + J_3 \gamma_{34} + J_5 \gamma_{56} + J_7 \gamma_{78} + \ldots  - i U_1 \gamma_{1234}
  \\ & ~~~- i U_3 \gamma_{3456} + U_5 \gamma_{5678} + U_7 \gamma_{789,10} + \ldots),
\end{align*}
where $\Omega = J_4$ refers to the strongest coupling. After rotating $\gamma_{2,3}$ and $\gamma_{6,7}$, we find
\begin{align*}
  F = & \exp( J_0 \gamma_{01} + \Omega \gamma_{45} + J_8 \gamma_{89} + \ldots -i U_0 \gamma_{01\tilde{2}\tilde{3}} -i U_{2} \gamma_{\tilde{2}\tilde{3}45}
  \\&~~~-i U_{4}\gamma_{45\tilde{6}\tilde{7}} -i U_{6}\gamma_{\tilde{6}\tilde{7}89} + \ldots ) \\
  \times &\exp(J_1 \gamma_{1\tilde{2}} + J_{1} J_{2} \gamma_{1\tilde{3}} +J_{3} \gamma_{\tilde{3}4} + J_{2} J_{3} \gamma_{\tilde{2}4} + J_{5} \gamma_{5\tilde{6}}  
  \\&~+J_{5} J_{6} \gamma_{5\tilde{7}} + J_{7}\gamma_{\tilde{7}8} + J_{7} J_{6} \gamma_{\tilde{6}8} + \ldots -i U_1 \gamma_{1 \tilde{2}\tilde{3}4}
  \\&~ -i U_3 \gamma_{\tilde{3}45\tilde{6}} -i U_3 J_{2} \gamma_{\tilde{2}45\tilde{6}} - i U_3 J_6 \gamma_{\tilde{3}45\tilde{7}} - i U_3 J_2 J_6 \gamma_{\tilde{2}45\tilde{7}} 
  \\&~-i U_5 \gamma_{5\tilde{6}\tilde{7}8} -i U_7 \gamma_{\tilde{7}89,10} - i U_7 J_6 \gamma_{\tilde{6}89,10}+ \ldots).
\end{align*}
We can therefore write the most general case as 
\begin{align}
F_0 &= e^{\Omega_{45} \gamma_4 \gamma_5}, \nonumber \\
iV &= J_{01} \gamma_{01} + J_{12} \gamma_{12} + J_{13} \gamma_{13} + J_{34} \gamma_{34} + J_{24} \gamma_{24} + J_{56} \gamma_{56} \nonumber \\
&+ J_{57} \gamma_{57} + J_{78}\gamma_{78} + J_{68} \gamma_{68} + J_{89} \gamma_{89}+ \ldots  \nonumber \\
&-i U_{0123} \gamma_{0123} -i U_{1234} \gamma_{1234} - i U_{2345} \gamma_{2345} -i U_{3456} \gamma_{3456} \nonumber \\
&-i U_{2456} \gamma_{2456} -i U_{3457} \gamma_{3457} - i U_{2457} \gamma_{2457} - i U_{4567} \gamma_{4567} \nonumber \\
&-i U_{5678} \gamma_{5678} -i U_{6789} \gamma_{6789} -i U_{789,10} \gamma_{789,10} \nonumber \\
&- i U_{689,10} \gamma_{689,10}+ \ldots,
\end{align}
where we have dropped the tilde's and relabelled the couplings more explicitly as $J_{ij} i \gamma_i \gamma_j$ and $U_{ijkl} \gamma_i \gamma_j \gamma_k \gamma_l$. Though tedious, these decimation rules are straightforward to compute. We find
\begin{align}
&\tilde J_{01} = J_{01}, 
 &\tilde J_{12} &= J_{12} + c \frac{J_{34} U_{1234}}{\tan \Omega_{45}}, \nonumber \\
&\tilde J_{13} = J_{13} - c \frac{J_{24} U_{1234}}{\tan \Omega_{45} }, 
~~~ &\tilde J_{23} &= c U_{2345}, \nonumber \\
&\tilde J_{26} = \frac{J_{24} J_{56}}{\tan \Omega_{45}} + c U_{2456},
~~~ &\tilde J_{27} &= \frac{J_{24}J_{57}}{\tan{ \Omega_{45}}} + c U_{2457}, \nonumber \\
&\tilde J_{36} = \frac{J_{34} J_{56}}{\tan \Omega_{45}} + c U_{3456},
~~~ &\tilde J_{37} &= \frac{J_{34} J_{57} }{\tan \Omega_{45}} + c U_{3457}, \nonumber \\ 
&\tilde J_{67} = -c U_{6789},
~~~ &\tilde J_{68} &= J_{68} - c \frac{J_{57} U_{5678}}{\tan \Omega_{45}}, \nonumber \\
&\tilde J_{78} = J_{78} + c \frac{J_{56} U_{5678}}{\tan \Omega_{45}},
~~~ &\tilde J_{89} &= J_{89}, \nonumber \\
~~~ &\tilde J_{9,10} = J_{9,10}, & & \nonumber \\
&\tilde U_{0123} = U_{0123},
~~~ &\tilde U_{1236} &= \frac{J_{56} U_{1234}}{\tan \Omega_{45}}, \nonumber \\
&\tilde U_{1237} = \frac{J_{57} U_{1234}}{\tan \Omega_{45}}, 
~~~ &\tilde U_{2678} &= \frac{J_{24} U_{5678}}{\tan \Omega_{45}}, \nonumber \\
&\tilde U_{3678} = \frac{J_{34} U_{5678}}{\tan \Omega_{45}}, 
~~~ &\tilde U_{689,10} &= U_{689,10},  \nonumber \\
&\tilde U_{789,10} = U_{789,10},
~~~ &\tilde W_{123678} &= \frac{U_{1234} U_{5678}}{\tan \Omega_{45}}.
\end{align}
Assuming $U \ll J$ as before, most of the interaction-induced dressing of the bilinears leads to irrelevant higher-order corrections. Interestingly, the interaction terms generate a new bilinear coupling between the two rotated sites, namely $\tilde J_{23} = c U_{2345}$ and $\tilde J_{67} = -c U_{6789}$, where the new bilinear is entirely due to the interaction term. However, note that this is still simply a bilinear coupling between the $0$ and $\pi$ quasi-energy chains, so again may be rotated away, and cannot affect the universality class of the transition.  

We have now computed how the couplings and quasi-energies are affected by the introduction of interaction terms. We find that the structure of the RG rules is very similar to the static case, except that since $\Omega$ is treated non-perturbatively, the denominators are $\tan \Omega$ instead of $\Omega$. As in the static case, then, interaction terms within each chain are irrelevant~\cite{PhysRevB.51.6411,PhysRevLett.69.534}, since they decay in strength as $\overline{\log(\Omega/U)} \sim \Gamma^{\phi}$ with $\Gamma \equiv \log \Omega_0 / \Omega$ the RG scale and $\phi=(1+\sqrt{5})/2$ the golden ratio, compared with $\overline{\log(J / \Omega) } \sim \Gamma$ for bilinear decimations. Tracking the interactions at the domain wall, we find that some interaction terms are generated that couple the two chains in the late stages of the RG; though by the above argument the intra-chain interactions are irrelevant, it is natural to wonder whether the inter-chain interactions are irrelevant as well. We show this by noting that our model maps explicitly onto a disordered XYZ model, $H_{XYZ} = \sum_i J_i^{XX} \sigma_i^x \sigma_{i+1}^x + J_i^{YY} \sigma_i^y \sigma_{i+1}^y + J_i^{ZZ} \sigma_i^z \sigma_{i+1}^z$. To derive the mapping, first apply a Jordan-Wigner transformation $\sigma_j^x = i \gamma_{2j} \gamma_{2j+1}$, $\sigma_j^y =  (\prod_{l < j} i \gamma_{2l} \gamma_{2l+1}) \gamma_{2j+1}$, $\sigma_j^z = (\prod_{l < j} i \gamma_{2l} \gamma_{2l+1}) \gamma_{2j}$. This implies $\sigma_j^x \sigma_{j+1}^x = - \gamma_{2j} \gamma_{2j+1} \gamma_{2j+2} \gamma_{2j+3}$, $\sigma_j^y \sigma_{j+1}^y = -i \gamma_{2j} \gamma_{2j+3}$, $\sigma_j^z \sigma_{j+1}^z = i \gamma_{2i+1} \gamma_{2i+2}$. This gives $H_{XYZ} = \sum_i J_i^{ZZ} i \gamma_{2i+1} \gamma_{2i+2} - J_i^{YY} i \gamma_{2i} \gamma_{2i+3} - J_i^{XX} \gamma_{2i} \gamma_{2i+1} \gamma_{2i+2} \gamma_{2i+3}$. This is manifestly of the form of two disordered Majorana chains coupled by density-density interactions, i.e. our model. Interactions in the XYZ model were studied and found to be irrelevant by Slagle et. al.~\cite{PhysRevB.94.014205}; hence, we conclude that these interactions also cannot change the universal critical physics of the free model above.

Note that this argument shows that interactions are irrelevant at the infinite-randomness fixed point. For ground states, it is possible to show that even weak disorder ultimately flows to this infinite-randomness fixed point, even in the presence of interactions. For excited states (i.e., RSRG-X) and for Floquet systems, however, there remains the possibility that rare many-body resonances can disrupt the  flow to infinite randomness, resulting ultimately in thermalization~\cite{PhysRevB.95.155129}. Such resonances, which are enabled by interactions, are not captured by the RSRG and therefore not ruled out by our treatment above which relied on proximity to the infinite-randomness fixed point. Therefore, we must always leave open the possibility that the ultimate fate of the critical/multi-critical system on the longest length and time scale is to thermalize to the infinite-temperature Gibbs state. This would be characterized by thermal correlations and volume-law ($\propto \ell$) entanglement scaling rather than the scaling discussed above. Nevertheless, for sufficiently strong disorder the dynamics on all reasonable (i.e., experimentally or numerically accessible) length and time scales will be controlled by the infinite-randomness fixed point, with a crossover to thermalization on exponentially long scales. Indeed, a definitive determination of which of these two scenarios occurs in the thermodynamic limit is outside the capability of current numerical simulations. We note that previous studies~\cite{khemani_prl_2016} of level statistics of the Floquet spectrum at criticality {for small system sizes and a given disorder strength} showed results intermediate between the Poisson statistics characteristic of localized systems and the circular orthogonal ensemble expected of thermalizing systems, {indicating that a systematic analysis of the dependence on disorder strength for larger system sizes would be needed.}

\section{Concluding Remarks\label{sec:discussion}}

In this paper, we have introduced a general method for performing real space renormalization for Floquet systems based on a generalization of the Schrieffer-Wolff transformation to Floquet unitaries. We have applied it to study the criticality in a paradigmatic model hosting several Floquet MBL phases and phase transitions -- the driven Ising model -- finding agreement with an earlier picture based on topological domain walls~\cite{Berdanier201805796}. Indeed, this calculation can be viewed as providing a microscopic derivation of the topological domain wall argument from a more mathematical perspective. 

This Floquet RSRG procedure can be readily applied to many one-dimensional Floquet MBL systems. For instance, a natural application would be to the periodically driven parafermion chain~\cite{PhysRevB.94.045127}, whose evolution operator has a similar structure to that of the driven Kitaev chain considered in this work. Importantly, this method does not depend on {\it a priori} knowledge of the phase structure of a given model, which can be deduced from the flow of the Floquet operator under renormalization. It is also applicable in cases where relative topological edge modes between different Floquet MBL phases either do not exist or are not known. In this vein, another natural problem to which this method may be fruitfully applied is to periodically driven anyon chains~\cite{PhysRevLett.114.217201}. {Finally, it is not always clear how to pick out a `strong' and `weak' piece of the Floquet unitary operator for every periodically driven system; for instance, when we drive the Ising model sinusoidally instead of in a piecewise fashion, it is no longer obvious how to determine $F_0$. We note that our previous topological arguments suggest that such a separation is indeed possible at late RG times~\cite{Berdanier201805796}. For now, we leave this factorization, as well as further applications and generalizations of our method, to future work. }

\section*{Acknowledgements}
W.B. acknowledges support from the Department of Defense (DoD) through the National Defense Science \& Engineering Graduate Fellowship (NDSEG) Program, and from the Hellman Foundation through a Hellman Graduate Fellowship. We also acknowledge support from Laboratory Directed Research and Development (LDRD) funding from Lawrence Berkeley National Laboratory, provided by the Director, Office of Science, of the U.S. Department of Energy under Contract No. DEAC02-05CH11231, the DoE Basic Energy Sciences (BES) TIMES initiative, and UTD Research Enhancement Funds (M.K.); travel support from the California Institute for Quantum Emulation (CAIQUE) via PRCA award CA-15- 327861 and the California NanoSystems Institute at the University of California, Santa Barbara (W.B. and S.A.P.); support from NSF Grant DMR-1455366 at the University of California, Irvine (S.A.P.); and 
US Department of Energy, Office of Science, Basic Energy Sciences, under Award No. DE-SC0019168 (R.V.)
The Mathematica package SNEG~\cite{ZITKO20112259} was used for several of the computations in this paper.

\bibliography{FloquetRSRG}
\bibliographystyle{apsrev4-1}

%

%
%
%
%

\end{document}